\documentclass[pre,twocolumn,superscriptaddress,showpacs]{revtex4}

\usepackage[dvips]{graphicx}
\usepackage{amsmath}

\newcommand{\D}{\ensuremath{\mathrm{d}}}
\newcommand{\Di}[1]{\ensuremath{\!\!\mathrm{d}#1\,}}
\newcommand{\Order}{\ensuremath{\mathrm{O}}}

\DeclareMathOperator{\erf}{erf}
\DeclareMathOperator{\erfc}{erfc}

\newcommand{\avg}[1]{\ensuremath{\langle #1 \rangle}}

\newcommand{\zero}{\ensuremath{\emptyset}}

\begin{document}

\title{Survival probability of a diffusing particle in the presence of
Poisson-distributed mobile traps}
\date{September 16, 2002; Revised: December 19, 2002}

\author{R.\ A.\ Blythe}
\affiliation{Department of Physics and Astronomy, University of 
Manchester, Manchester M13 9PL, U.K.}

\author{A.\ J.\ Bray}
\affiliation{Department of Physics and Astronomy, University of 
Manchester, Manchester M13 9PL, U.K.}
\affiliation{Laboratoire de Physique Quantique (UMR C5626 du CNRS), 
Universit\'e Paul Sabatier, 31062 Toulouse Cedex, France}
 
\begin{abstract}
The problem of a diffusing particle moving among diffusing traps 
is analyzed in general space dimension $d$. We consider the case 
where the traps are initially randomly distributed in space, with 
uniform density $\rho$, and derive upper and lower bounds for 
the probability $Q(t)$ (averaged over all particle and trap 
trajectories) that the particle survives up to time $t$. We 
show that, for $1 \le d \le 2$, the bounds converge asymptotically 
to give $Q(t) \sim\exp(-\lambda_d t^{d/2})$ for $1 \le d < 2$, 
where $\lambda_d = (2/\pi d) \sin(\pi d/2)(4\pi D)^{d/2}\rho$ and 
$D$ is the diffusion constant of the traps, and that 
$Q(t) \sim \exp(-4\pi\rho Dt/\ln t)$ for $d=2$. For $d>2$ bounds 
can still be derived, but they no longer converge for large $t$. 
For $1 \le d \le 2$, these asymptotic form are independent of 
the diffusion constant of the particle. The results are compared 
with simulation results obtained using a new algorithm [V. Mehra 
and P. Grassberger, Phys. Rev. E {\bf 65}, 050101 (2002)] which 
is described in detail. Deviations from the predicted asymptotic 
forms are found to be large even for very small values of $Q(t)$, 
indicating slowly decaying corrections whose form is consistent 
with the bounds.  We also present results in $d=1$ for the case 
where the trap densities on either side of the particle are different. 
For this case we can still obtain exact bounds but they no longer 
converge.  
\end{abstract}

\pacs{05.40.-a, 02.50.Ey, 82.20.-w}

\maketitle

% -----------------------------------------------------------------------------
\section{Introduction}

Reaction-diffusion processes represent a large and important class of
systems with nonequilibrium dynamics.  From a fundamental physical
viewpoint, the interest in these systems lies in the fact that the
concentration of reactants is governed, in general, by irreversible
reaction events that depend on the spatial distribution of particles
rather than through equilibrium fluctuations controlled by a chemical
potential.  Such model systems have a range of applications, most
notably to chemical kinetics \cite{Benson60,Rice85} but also to
interfacial growth \cite{KS91}, domain coarsening \cite{Bray94,DZ96} 
and aggregation \cite{Spouge88}.

The most intensively studied reactions are single-species annihilation
($A+A\to\zero$) and coalescence ($A+A\to A$) as well as two-species
annihilation ($A+B\to\zero$)---see, e.g.,
\cite{Redner97,benAvraham97,bAH00} for reviews.  In this paper we
focus on the two-species problem. It is known to exhibit two different
classes of long-time behavior depending on whether the initial
concentrations of $A$ and $B$ particles are equal or not.  (As an
aside, we note that a similar dependence on the initial condition also
holds for the $A+A\to\zero$ reaction when the reactant motion is
deterministic rather than diffusive \cite{EF85,BEK00}).  The reason
for this is that when the initial densities of $A$ and $B$ particles
are the same, they remain so for all times, whereas if, say, the
initial density of $A$ particles $\rho_A(0)$ is less than that of the
$B$ particles $\rho_B(0)$, the ratio $\rho_A(0)/\rho_B(0) \to 0$ as $t
\to \infty$ and at late times one has a few, isolated $A$ particles
diffusing in a background of $B$ particles.

The case of equal initial densities is well understood, and results
similar to those for the $A+A\to\zero$ with diffusive particle motion
have been obtained \cite{TW83,LC95}.  In low dimensions, here $d<4$,
fluctuation effects are important and one finds a density decay
$\rho_A(t)=\rho_B(t) \sim t^{-d/4}$ in this \textit{diffusion-limited}
regime.  Above the critical dimension $d>d_c=4$ one finds that the
mean-field result $\rho_{A,B}(t) \sim 1/t$ applies.  This result also
holds for the $A+A\to\zero$ process above its critical dimension $d_c=2$.

By contrast, the density decay forms for the $A+B\to\zero$ process
when the initial densities $\rho_A(t)$ and $\rho_B(t)$ are not equal
are less well understood.  In fact, since the exposition of the
process as a model of monopole-antimonopole annihilation in the early
universe nearly twenty years ago \cite{TW83}, only a few results are
known exactly.  Most notably, Bramson and Lebowitz \cite{BL88} proved
rigorously that, at large times, the density of the minority species
(which we will take to be the $A$ particles) behaves as
\begin{equation}
\label{BLresults}
\rho_A(t) \sim \left\{ \begin{array}{ll}
\exp( - \lambda_d t^{d/2} ) & d < 2 \\
\exp( - \lambda_2 \ln t/t ) & d = 2 \\
\exp( - \lambda_d t ) & d > 2
\end{array} \right.
\end{equation}
revealing $d=2$ to be critical in this case.  To the best of our
knowledge, no predictions for the constants $\lambda_d$ were given
until recently \cite{BB02a}.  Furthermore, there has been no 
convincing numerical verification of the predicted decay even in 
one-dimension, despite the development of sophisticated simulation 
techniques \cite{MG02} that allow the probing of extremely small
densities that emerge at large times.  In this paper, we expand on the
bounding arguments reported in \cite{BB02a} that give rise to precise
values of $\lambda_d$ for $d \le 2$.  We also present a detailed
description of the simulation algorithm introduced in \cite{MG02} and
extend it to test our bounding arguments and understand the approach
to the asymptopia described by Eq.\ (\ref{BLresults}).

As noted above, the late-time regime is characterized by a few
isolated $A$ particles diffusing in a sea of $B$ particles.  Thus it
is appropriate to consider the extreme case of a single $A$ particle
in a sea of $B$ particles that has a uniform (Poisson) density.  In
this case, the quantity $\rho_A(t)$ is just the \textit{survival
probability} of the $A$ particle.  Furthermore, if the diffusion
constants of the $A$ and $B$ particles are the same, one can also view
$\rho_A(t)$ as the fraction of particles that have not met any other
particles.  Thus the reaction $A+B\to \emptyset$ in the limit of a low
density of $A$ particles has been discussed under the guises of
\textit{uninfected walkers} \cite{OB02} in which random walkers infect
each other on contact, \textit{diffusion in the presence of traps}
\cite{DV75,MG02} in which the $B$ particles are considered as traps
for the $A$ particles, and \textit{predator-prey models} \cite{RK99}
in which one asks for the survival of a prey (the $A$ particle) being
`chased' by diffusing predators (the $B$ particles).  To avoid
confusion, we shall adopt only the trapping terminology in our
discussion.

In this work, we show how the survival probability of a diffusing
particle in the presence of mobile traps can be understood in terms of
the \textit{target annihilation problem} \cite{BZK84,Tachiya83,BO87}
(or first passage problem \cite{Redner01}) where one asks for the
probability that none of the traps has entered a particular region
(target) in the $d$-dimensional space.  In turn, the asymptotics of
the target annihilation problem are intimately related to the
\textit{recurrence} or \textit{transience} of diffusion in various
dimensions.  A process is said to be recurrent if the probability of
returning to the initial configuration is unity: in the context of
diffusion, this implies that with probability one a walker will visit
a particular point in space infinitely often.  It is well known (see,
e.g, \cite{Weiss94,Redner01}) that diffusion is recurrent in
dimensions $d \le 2$, whereas in more than two dimensions it is
transient (i.e.\ the return probability is less than one). It is
precisely this property of diffusion that gives rise to the critical
dimension of two for the trapping reaction and hence the asymptotic
results (\ref{BLresults}) for the $A+B\to\zero$ process.

The principal result of the paper is the determination of the constants  
$\lambda_d$ in Eq.\ (\ref{BLresults}) for $d \le 2$, and the derivation of 
upper and lower bounds for $d=3$. A striking feature of the results 
is that, for $d \le 2$, the value of $\lambda_d$ is independent of the 
diffusion constant of the $A$ particle. 

We begin in the next section of this paper by defining the trapping
reaction model.  Then, in section~\ref{1danalysis} we present in detail
our analysis of the one-dimensional case, testing our predictions in
section~\ref{numerics} where we discuss how the model may be simulated
efficiently. In section~\ref{generaldanalysis}, we show how the method
used to treat the one-dimensional case can be extended to general
dimensions $d>1$. Only when the underlying diffusion process is
recurrent (i.e.\ for $d\le2$) do our upper and lower bounds converge
asymptotically to give exact predictions for $\lambda_d$.  Finally, in
section \ref{summary}, we present a discussion and summary of the
results.

\section{Definition of the model}
\label{moddef}

The trapping reaction model we consider is defined as follows.  At
time $t=0$ a particle is placed at the origin of a $d$-dimensional
coordinate system.  Surrounding this particle is a uniform sea of
traps whose initial positions $\vec{x}_i$ are chosen independently.
This initial condition ensures that the distribution of traps is
Poisson, i.e.\ the probability that a volume $V$ contains $N$ traps is
$[(\rho V)^{N}/N!] \exp(-\rho V)$ in which $\rho$ is the mean number of
traps per unit volume.

The dynamics of the particle and traps can be expressed using the
Langevin equation
\begin{equation}
\dot{x}_i^{\alpha} = \eta_i^\alpha(t)
\end{equation}
in which the subscript $i=0$ denotes the particle, $i>1$ one of the
traps and the superscript $\alpha$ indicates a component of the
position vector $\vec{x}_i$.  The noise $\eta_i^\alpha(t)$ is a
Gaussian white noise with zero mean and correlator
\begin{equation}
\avg{\eta_i^\alpha(t) \eta_j^\beta(t^\prime)} = 2 D_i \delta_{ij}
\delta_{\alpha \beta} \delta(t-t^\prime) \;.
\end{equation}
We take all the traps to have a diffusion constant $D$ and the
particle to have a diffusion constant $D^\prime$.  Hence
$D_0=D^\prime$ and $D_i = D$ for $i>0$.  The quantity of
interest in this model is the probability $Q(t)$, averaged over all 
initial conditions and realizations of the random walks, that the 
particle has not yet met any of the diffusing traps.  

\section{Analytical results in one dimension}
\label{1danalysis}

For clarity, we restrict ourselves initially to the case $d=1$.
Later, in section~\ref{generaldanalysis} we will explain how the
arguments presented in detail here can be generalized to higher
dimensions.  We begin with a description of the target annihilation
problem before moving on to discuss how it applies to the more general
problem of a particle's survival in a sea of diffusing traps. The
target annihilation problem can be solved exactly for any $d$
\cite{BZK84,Tachiya83,BO87}.  The asymptotic form of the solution, and
the leading corrections to it (for $d>1$), play a central role in our
bounding arguments.  To establish the notation and to make our
presentation self-contained, we present in this paper a brief
derivation of the main results as a prelude to deriving the bounds.

\subsection{The target annihilation problem}
\label{1dtarget}

Consider a one-dimensional line containing a target of length $2l$
centered on the origin (i.e.\ lying between $x=-l$ and $x=l$).
We wish to calculate the probability $Q_T(t)$ that none of the
diffusing traps initially placed outside this region has hit the
target by a time $t$.  This quantity can be calculated if one knows
the probability $Q_1(t|y)$ that a trap initially at position $y$ has
not yet entered the target region.  Since the target is static and
each trap executes independent diffusion, we can simply multiply the
probabilities for each individual trap together and average over all
possible initial positions to find $Q_T(t)$.

Let us consider then a trap that has its initial position to the right
of the target, i.e.\ $y>l$.  The probability $Q_1(t|y)$ that the
trap has not reached the target satisfies the backward Fokker-Planck
equation
\begin{equation}
\label{bfpe1}
\frac{\partial Q_1(t|y)}{\partial t} = D \frac{\partial^2
Q_1(t|y)}{\partial y^2}
\end{equation}
with the boundary conditions $Q_1(t|l) = 0$, $Q_1(0|y)=1$ if
$y>l$ and $Q_1(t|\infty) = 1$.  These express the facts that the
probability that the target has been reached if the trap started at
$y=l$ is one, that it is reached in zero time from $y>l$ is zero
and that it is reached from infinity \textit{in a finite time} is zero
respectively.  The solution to (\ref{bfpe1}) that satisfies these
boundary conditions is
\begin{equation}
Q_1(t|y) = \erf \left( \frac{y-l}{\sqrt{4Dt}} \right)
\end{equation}
in which $\erf(x)$ is the error function.

Instead of a single trap to the right of the target, consider $N$
independently diffusing traps, each initially placed at random in the
interval $y_i \in [l, l+L]$.  Then, the probability that none of
the traps has reached the target by time $t$ is
\begin{equation}
Q_N(t) = \prod_{i=1}^{N} \frac{1}{L} \int_{l}^{l+L} \Di{y_i} \erf
\left( \frac{y_i-l}{\sqrt{4Dt}} \right) \;.
\end{equation}
It is convenient now to rewrite the error function in terms of the
complementary error function, $\erf(x) = 1 - \erfc(x)$.  Then one has
\begin{equation}
Q_N(t) = \left[ 1 - \frac{1}{L} \int_{l}^{l+L} \Di{y} \erfc
\left( \frac{y-l}{\sqrt{4Dt}} \right)  \right]^N \;.
\end{equation}
Since we wish to consider an infinite sea of traps, we take $N = \rho
L$ and then the limit $L \to \infty$ holding $\rho$, the density of
traps, fixed.  This yields
\begin{eqnarray}
\label{1doneside}
Q_\infty(t) &=& \lim_{L \to \infty} \left[ 1 - \frac{1}{L}
\int_l^{l+L} \Di{y} \erfc \left( \frac{y-l}{\sqrt{4Dt}}
\right) \right]^{\rho L} \nonumber\\
&=& \exp \left( -\frac{2 \rho \sqrt{Dt}}{\sqrt{\pi}} \right) \;.
\end{eqnarray}
This gives the probability that no traps initially positioned on one
side of the target have reached the target by time $t$.  Since we have
in mind a target surrounded on both sides by traps, and that the
motion on each side is independent, we obtain the probability that the
target has not been annihilated by a trap by squaring
(\ref{1doneside}).  That is,
\begin{equation}
\label{QT1d}
Q_T(t) = \exp \left( -\frac{4 \rho \sqrt{Dt}}{\sqrt{\pi}} \right) \;.
\end{equation}
Note that the size of the one-dimensional target $l$ does not
appear in this exact expression for its survival probability.
Later, in section~\ref{generaldanalysis}, we will find that at
suitably large times, the size of the target is unimportant for all
$d<2$ (where diffusion is recurrent).

\subsection{Bounding argument for a diffusing particle in the presence of
mobile traps}
\label{1dbounds}

We now discuss how to construct upper and lower bounds on the
particle's survival probability $Q(t)$ using the result for the target
annihilation problem (\ref{QT1d}) in one dimension.  We claim that, on
average, a particle surrounded by a uniform, isotropic distribution of
traps survives longer if it is stationary than if it is allowed to
diffuse.  We are currently unable to prove this statement rigorously,
although it is supported by intuition and numerical data (see
section~\ref{numerics} below).  We also note that when we say ``on
average'' we mean ``after averaging over all possible initial trap
positions and trajectories of both particle and traps''.

If this claim is accepted, we obtain an upper bound $Q_U(t)$ on the
particle's survival probability from $(\ref{QT1d})$ by noting that
requiring the particle to remain stationary is equivalent to having a
target region of size $l = 0$.  Then we immediately have that
\begin{equation}
\label{QU1d}
Q(t) \le Q_U(t) = \exp \left( -\frac{4 \rho \sqrt{Dt}}{\sqrt{\pi}}
\right) \;.
\end{equation}

\begin{figure}
\includegraphics[width=\linewidth]{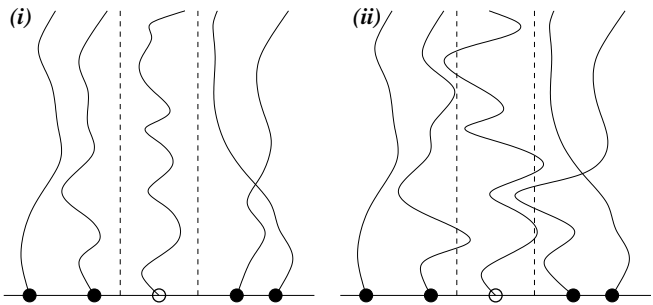}
\caption{\label{subset}Two walker trajectories (space-time plots, 
with $t=0$ at the bottom) for 
which the particle (unfilled) survives contact with a trap.  
Trajectories of type (i) have the property that the particle remains 
inside a notional box, and the traps outside.  This forms a subset of 
the entire class of surviving trajectories, which includes paths 
of type (ii) in which the particle leaves the box and the traps enter 
but nevertheless no particle-trap contact occurs.}
\end{figure}

To derive a lower bound on the survival probability $Q(t)$ we
introduce a notional box of size $l$ centered on the origin.  If we
ask for the particle to remain inside this box until a time $t$, and
for all the traps to remain outside it, the traps and particle may
never meet and hence the particle survives until time $t$.  There are,
of course, other trajectories for which the particle survives, and so
those just described form a subset of all possible surviving
trajectories---see Fig.~\ref{subset}.  Hence the probability
that the particle remains within the box and traps outside
is a lower bound $Q_L(t)$ on $Q(t)$.

There are three independent contributions to this bound: (i) the
probability that there are initially no traps in the box of size
$l$; (ii) the probability that no traps enter the box up to time
$t$; and (iii) the probability that the particle has not left the box
up to time $t$.  The first two contributions are easily obtained.
From the definition of the Poisson distribution, we have that the
probability the box initially contains no traps is $\exp(-\rho l)$.
Secondly, the probability that no traps enter the box is independent
of the box size and is given by (\ref{QT1d}).  The third contribution,
the probability that the particle remains inside the box, is obtained
as follows.

Since the system is translationally invariant, we can just as easily
consider a particle initially sandwiched between absorbing boundaries
at $x=0$ and $x=l$.  The probability $Q_P(t|y)$ that the particle
starting at $y=l/2$ has not crossed the absorbing boundaries
satisfies a backward Fokker-Planck equation
\begin{equation}
\frac{\partial Q_P(t|y)}{\partial t} = D^\prime \frac{\partial^2
Q_P(t|y)}{\partial y^2}
\end{equation}
subject to the absorbing boundary conditions $Q_P(t|0)=Q_P(t|l)=0$
and the initial condition $Q_P(t|y) = 1$ for $0 < y < l$.  The
general solution to this equation that satisfies the absorbing
boundary conditions is obtained by separating the time and space
variables in the usual way.  One obtains the Fourier sine series
\begin{equation}
\label{QPsin}
Q_P(t|y) = \sum_{k=1}^{\infty} a_k \exp\left(- \frac{k^2 \pi^2
D^\prime t}{l^2} \right) \sin\left( \frac{k \pi y}{l} \right) \;.
\end{equation}
The expansion coefficients $a_k$ are fixed through the initial
condition.  Using the orthogonality of the sine functions one finds 
\begin{equation}
a_k = \left\{ \begin{array}{cl}
\displaystyle \frac{4}{\pi k} & \mbox{$k$ odd} \\[2ex]
0 & \mbox{$k$ even}
\end{array} \right. \;.
\end{equation}

For the purposes of the present calculation, we need consider only
the long-time form of $Q_P(t|y)$ for a particle that starts at $y=
l/2$.  Thus we keep only the longest-lived ($k=1$) mode in the
expansion (\ref{QPsin}) to find
\begin{equation}
Q_P(t|l/2) \sim \frac{4}{\pi} \exp \left( - \frac{\pi^2 D^\prime
t}{l^2} \right) \;.
\end{equation}
Including this along with the contributions to the lower bound
$Q_L(t)$ on the diffusing particle's survival probability $Q(t)$
discussed above, we have
\begin{equation}
\label{QLloose}
Q(t) \ge Q_L(t) = \frac{4}{\pi} \exp\left( -\frac{\pi^2 D^\prime t}{l^2}
-\rho l -\frac{4\rho\sqrt{Dt}}{\sqrt{\pi}} \right)
\end{equation}
once the time $t$ is sufficiently large.  Note that this provides a bound
for a particular box size $l$.  Since the box is an artificial
construct, we can choose its size so that the lower bound is maximized
at a particular (predetermined) time $t^\ast$.  One finds that the
corresponding box size is $l^\ast = (2\pi^2 D^\prime t^\ast/
\rho)^{1/3}$.  Using this box size in (\ref{QLloose}) we find the
largest lower bound is given by
\begin{equation}
\label{QL1d}
Q_L(t) = \frac{4}{\pi} \exp\left( -\frac{4\rho\sqrt{Dt}}{\sqrt{\pi}} -
3 \left[ \frac{\pi^2 \rho^2 D^\prime t}{4} \right]^{1/3} \right) \;.
\end{equation}

Combining this lower bound with the upper bound $Q_U(t)$ of
Eq.\ (\ref{QU1d}) we find
\begin{equation}
\frac{4}{\sqrt{\pi}} \le - \frac{\ln Q(t)}{(\rho^2 Dt)^{1/2}} \le
\frac{4}{\sqrt{\pi}} + 3 \left(\frac{\pi}{2}\right)^{2/3}
\frac{(D^\prime/D)^{1/3}}{(\rho^2 Dt)^{1/6}} \;.
\end{equation}
This implies that the constant $\lambda_1$ in the expressions of
Bramson and Lebowitz (\ref{BLresults}) is precisely determined as
\begin{equation}
\lambda_1 = - \lim_{t\to\infty} \frac{\ln Q(t)}{\sqrt{t}}
= 4 \rho \sqrt{\frac{D}{\pi}} \;.
\label{lambda1}
\end{equation}
Note that this constant depends only on the density and diffusion
constant of the traps, and is independent of the diffusion constant 
of the particle.

\subsection{Extensions to the basic trapping reaction model}
\label{extensions}

It is straightforward to incorporate two generalizations of the
one-dimensional trapping model defined in section \ref{moddef} into
the bounding arguments discussed above.  The first of these is to
allow the traps to the left and right of the origin at time $0$ to
have different densities.  We denote the larger (respectively,
smaller) of these densities as $\rho_+$ ($\rho_-$) and their average
as $\bar{\rho} = \frac{1}{2}(\rho_++\rho_-)$.  Additionally we shall
place $n$ particles at the origin at time $0$ and study the
probability that \textit{all} survive until a time $t$.

To obtain an upper bound on the survival probability, we note that the
survival probability of the particles can only increase (or remain
constant) as either $\rho_+$ or $\rho_-$ is decreased.  Hence the
survival probability for the case of unequal densities is bounded from
above by the case where the density of traps is on both sides equal to 
$\rho_-$.  For the case of a single diffusing particle, we argued
above that an upper bound on its survival probability is found by
setting its diffusion constant $D^\prime$ to $0$.  Clearly, if
$D^\prime=0$ the number of particles at the origin is irrelevant, and
so an upper bound on $Q(t)$ is given by Eq.\ (\ref{QU1d}) with
$\rho=\rho_-$, i.e.,
\begin{equation}
\label{nupper}
Q(t) \le Q_U(t) = \exp\left( -\frac{4\rho_{-}\sqrt{Dt}}{\sqrt{\pi}}
\right) \;.
\end{equation}

To obtain a lower bound on the particles' survival probability we
once again introduce a notional box, inside which all the particles
must remain and no traps may enter until time $t$.  This time,
however, we respect the asymmetry of the problem by allowing the box
to extend a distance $l_-$ into the low-density region of traps and
$l_+$ into the high-density region.  We will again seek to maximize
the lower bound by varying $l_-$ and $l_+$.

A lower bound $Q_L(t)$ is obtained using an argument analogous to
that leading to Eq.\ (\ref{QL1d}).  Considering once again late times,
we find
\begin{widetext}
\begin{equation}
\label{nlower1}
Q_L(t) \propto \exp\left( - \frac{n \pi^2 D^\prime t}{(l_- +
l_+)^2} - (\rho_-l_- + \rho_+l_+) - \frac{2\rho_-
\sqrt{Dt}}{\sqrt{\pi}} - \frac{2\rho_+ \sqrt{Dt}}{\sqrt{\pi}} \right)
\;.
\end{equation}
\end{widetext}
The number of particles $n$ enters into this expression through the
fact that the probability for \textit{all} of the $n$ particles to
remain inside the box of size $l = l_- + l_+$ is simply the
$n^{\mathrm{th}}$ power of the corresponding probability for a single
particle.

The maximal lower bound for a prescribed time $t^\ast$ is obtained
from (\ref{nlower1}) by setting $l_+^\ast$ to zero (thus discounting
particle trajectories that enter the high-density region) and putting
$l_-^\ast = (2n\pi^2 D^\prime t^\ast / \rho_-)^{1/3}$.  Then
\begin{equation}
Q(t) \ge Q_L(t) \propto \exp\left( - \frac{4 \bar{\rho}
\sqrt{Dt}}{\sqrt{\pi}} - 3 \left[ \frac{n \pi^2 \rho_-^2 D^\prime
t}{4} \right]^{1/3} \right) \;.
\end{equation}
Along with the upper bound (\ref{nupper}) we find that
%\begin{equation}
\begin{multline}
\label{genbounds}
\frac{4}{\sqrt{\pi}} \le - \frac{\ln Q}{(\rho_-^2 Dt)^{1/2}} \le
\frac{4}{\sqrt{\pi}} \frac{\bar{\rho}}{\rho_-} + {} \\
3 \left(\frac{n \pi^2 D^\prime}{4D}\right)^{1/3} \frac{1}{(\rho_-^2
Dt)^{1/6}} \;.
\end{multline}
%\end{equation}
Note that, except for the case where $\bar{\rho}=\rho_-$ (which
implies $\rho_- = \rho_+$) these two bounds do not converge and so we
cannot make a precise prediction for $\lambda$ when the trap densities 
are unequal. For the case $\rho_-=\rho_+$, however, the bounds converge
to $4/\sqrt{\pi}$, independent of the number of particles $n$.

\section{Simulation algorithm and results}
\label{numerics}

A sophisticated algorithm for simulating the trapping reaction in
discrete space and time and with a Poisson distribution of traps was
recently introduced \cite{MG02}.  The beauty of the algorithm is that
it admits (numerically) exact calculation of the survival probability
for an arbitrarily long, but fixed, trajectory of the particle.  As
will be discussed below, the algorithm takes into account all possible
paths of the traps, as long as their initial distribution is Poisson.
In order to obtain an estimate of the particle survival probability,
it is necessary to iterate the algorithm over a sequence of particle
paths.  We now discuss this algorithm in detail.

\subsection{An efficient simulation algorithm}

In order to simulate the trapping reaction model in one dimension, we
construct a discretized version in which each walker follows a path
$x(t)$ that has $x(t+1)-x(t) = \pm 1$.  Since all hops to the left or
right occur in parallel, we must ensure that the initial coordinates
of all the walkers are even integers so that no two walkers are able
to hop over each other.

As a starting point in understanding the simulation algorithm,
consider a system comprising the particle, whose trajectory $x_0(t)$
is predetermined, and a single trap, whose trajectory $x_1(t)$ is
stochastic given some initial condition $x_1(0)=y_1$.  The probability
$P_1(x,t)$ of finding the trap at site $x$ after time $t$, given that it
has not absorbed the particle, satisfies the equation
\begin{equation}
\label{discretede}
P_1(x, t+1) = \frac{1}{2} \left[ P_1(x-1, t) + P_1(x+1, t) \right]
\end{equation}
subject to the initial condition $P(x, 0) = \delta_{x,y_1}$ and the
moving absorbing boundary condition $P(x_0(t), t) = 0$.  Note that
(\ref{discretede}) is the discrete analogue of the diffusion
(Fokker-Planck) equation
\begin{equation}
\frac{\partial P_1(x,t)}{\partial t} = D \frac{\partial^2
P_1(x,t)}{\partial x^2} \;.
\end{equation}
By Taylor expanding (\ref{discretede}) we find the diffusion constant
of both particle and trap to be $D= D^\prime = \frac{1}{2}$.

The solution of the diffusion equation with an arbitrary moving
absorbing boundary at $x_0(t)$ is not known analytically.  One can
obtain it numerically, however, by iterating the following two steps
over $t^\prime = 1, 2, \ldots, t$.
\begin{enumerate}
\item Construct the probability distribution of the trap's position
using the equation $P_1(x, t^\prime) = \frac{1}{2} \left[ P_1(x-1,
t^\prime-1) + P_1(x+1, t^\prime-1) \right]$.
\item Enforce the absorbing boundary condition by subsequently
setting $P_1(x_0(t^\prime), t^\prime) = 0$.
\end{enumerate}

In the simulation, we wish to consider not just a single trap, but a
Poisson distribution of traps.  This can be achieved as follows.  Let
$P_n(x, t)$ be the probability that there are $n$ traps on lattice
site $x$ at time $t$.  We shall assume that this distribution is
Poisson, i.e.,
\begin{equation}
P_n(x,t) = \frac{[c(x,t)]^n}{n!} \exp[-c(x,t)]
\end{equation}
in which $c(x,t)$ is the mean number of traps at site $x$ and time
$t$.

Now, if each trap can hop with equal probability to the left or right
in one time step, we have
\begin{equation}
\label{masteresque}
P_n(x,t+1) = \sum_{m=0}^{n} W_m^+(x-1,t) W_{n-m}^-(x+1,t)
\end{equation}
in which $W_m^\pm(x,t)$ is the probability that $m$ particles hop from
site $x$ at time $t$ to $x\pm 1$ at time $t+1$.  This quantity is
given by
\begin{equation}
W_m^\pm(x,t) = \sum_{s=m}^{\infty} \frac{[c(x,t)]^s}{s!} \exp [-c(x,t)]
\binom{s}{m} \frac{1}{2^s} \;.
\end{equation}
Insertion of this expression into (\ref{masteresque}) and a little
algebra reveals that
\begin{equation}
P_n(x,t+1) = \frac{[\bar{c}(x,t)]^n}{n!} \exp [ - \bar{c}(x,t) ]
\end{equation}
in which
\begin{equation}
\bar{c}(x,t) = \frac{c(x-1,t)+c(x+1,t)}{2} \;.
\end{equation}
That is, if the distribution of traps at time $t$ is Poisson the
distribution of traps at time $t+1$ is also Poisson, with the mean
occupation number at each site obeying the discrete diffusion equation
\begin{equation}
c(x,t+1) = \frac{1}{2} \left[ c(x-1,t) + c(x+1,t) \right] \;.
\end{equation}

As with the case of the single trap described above, we wish to
determine the probability distribution of traps given that the
particle following the predetermined path $x_0(t)$ has not been
absorbed until a time $t$.  We must therefore have at each time step
$P_n(x_0(t), t) = \delta_{n,0}$ which can be achieved by enforcing the
boundary condition $c(x_0(t),t) = 0$.  Thus we can evolve the mean
occupation numbers for the Poisson distributed sea of traps in exactly
the same way as for the single-trap distribution function described
above (albeit with a different initial condition, to be described
shortly).

In the simulations, we wish to calculate the probability that the
particle has survived until time $t$.  To obtain an expression for
this, consider a particular distribution of traps described by the
function $c(x,t)$.  The probability that site $x_0$ contains no traps
is just $\exp(-c[x_0,t])$ and so $Q(t+1) = Q(t) \exp(-c[x_0,t])$
where the value of $c(x_0,t)$ used is that obtained after the
diffusion step, but before enforcing the boundary condition $c(x_0,t)
= 0$.

We now give a step-by-step explanation of the algorithm for
calculating the particle survival probability for a predetermined
particle path $x_0(t)$.  One begins by setting up the trap
concentration as follows:
\begin{equation}
c(x,t_0) = \left\{ \begin{array}{ll} 2 \rho_L & x < x_0(0) \\ 0 & x =
x_0(0) \\ 2 \rho_R & x > x_0(0) \end{array} \right. 
\end{equation}
in which $\rho_L$ and $\rho_R$ are the equivalent continuum densities
to the left and right of the particle, as used in section
\ref{extensions}.  The factor of $2$ emerges because that is the
effective lattice spacing in the discrete model.  We also set $Q(0)
= 1$ (i.e., we assume there are no traps at the origin to begin
with).  Then, for each time $t^\prime = 1,2, \ldots, t$ we perform the
following steps:
\begin{enumerate}
\item The trap concentration variables are evolved using
$c(x,t^\prime) = \frac{1}{2}\left[ c(x-1,t^\prime-1) + c(x+1,t^\prime-1)
\right]$.
\item The cumulative particle survival probability is calculated using
$Q(t^\prime) = Q(t^\prime-1) \exp \left[-c(x_0(t^\prime),t^\prime) \right]$.
\item The boundary condition is enforced by setting
$c(x_0(t^\prime),t^\prime)=0$.
\end{enumerate}
Note that this algorithm can be run for paths of arbitrary length 
and that, at a particular time $t^\prime$, the trap density at
positions $x < x_0(0) - t^\prime$ and $x > x_0(0) + t^\prime$
is uniform.  Hence at each time step, one need deal only with 
$t^\prime+1$ concentration variables to simulate the infinite system.

Using the above algorithm, one obtains the survival probability for a
particle following a particular path $x_0(t)$.  To reach an estimate
of the particle survival probability averaged over all paths, it is
most efficient to perform Monte Carlo sampling.  That is, one
generates a binomial random walk by choosing the particle displacement
$x_0(t^\prime)-x_0(t^\prime-1) = \{-1,1\}$ with equal probability.
Then, one estimates the mean particle survival probability as
\begin{equation}
Q(t) \approx \frac{1}{N} \sum_{k=1}^{N} Q^{(k)}(t)
\end{equation}
in which $Q^{(k)}$ is the value of the survival probability for the
$k^{\mathrm{th}}$ random walk.  One can, of course, estimate other
quantities, such as the mean and variance of the particle's
displacement.  Also, if one is interested only in the short-time
behavior, one can obtain the particle survival probability for each
possible path.  We should also note that the one-dimensional algorithm
described here generalizes straightforwardly to higher (integer)
dimensions.

\subsection{Numerical results}

We first investigate the entire set of short particle paths in order
to get a feel for those that give rise to the greatest probability of
survival.  For each time $t \le 28$ we found that the paths which have
the greatest survival probability are those with the smallest width
(defined as the distance between the extrema of the path), i.e.\ the
sequences $x(t) = (0,1,0,1,0,\ldots)$ and $x(t) =
(0,-1,0,-1,0,\ldots)$.  This result gives support to the supposition
in section~\ref{1danalysis} that staying still (i.e.\ a diffusion
constant $D^\prime=0$) gives rise to the greatest chance of survival.
We also established this to be case for two-dimensional walks up to a
time $t = 12$.

\begin{figure}
\includegraphics[width=\linewidth]{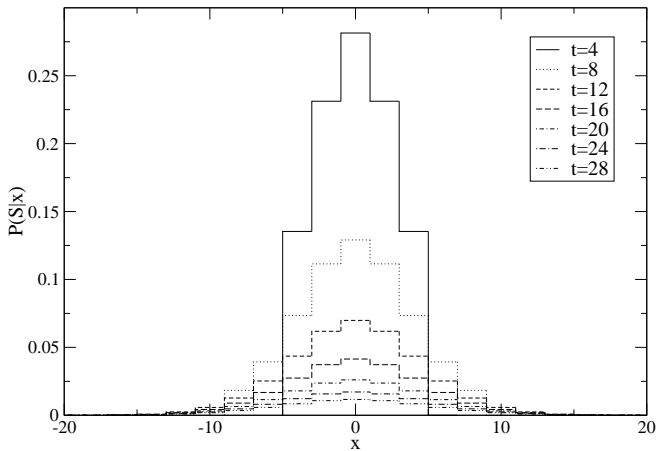}
\caption{\label{histogram}Survival probability $P(S,t|x)$ given that
the particle is at site $x$ at early times and with
$\rho_L=\rho_R=0.5$.}
\end{figure}

It is a simple matter to use the algorithm presented above to find the
probability $P(x,t|S)$ for the particle to be at coordinate $x$ after
time $t$ given that it has survived.  Then, an application of Bayes'
theorem yields the more telling quantity $P(S,t|x)$, i.e.\ the
probability that the particle has survived to time $t$ given that it
ends at coordinate $x$.  The resulting data are plotted in
Fig.~\ref{histogram} and one sees quite clearly that that the particle
is most likely to survive if it is at the origin, at least for times
$t\le28$.  This figure provides further weight to our assertion that
staying still is the best particle survival strategy.

As stated in the previous section, one can obtain estimates of various
quantities at later times if one performs Monte Carlo sampling over
particle paths. In fact, we produced histograms of $P(S,t|x)$ this way
and obtained data very similar to those shown in Fig.~\ref{histogram}
(except with poorer statistics).  Hence we do not present them here.
Instead we concentrate on the survival probabilities for a range of
trap densities to compare with the bounds given by (\ref{genbounds}).

\begin{figure}
\includegraphics[width=\linewidth]{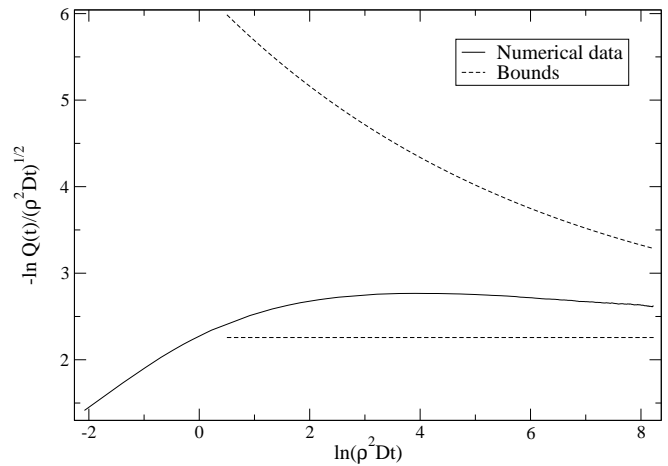}
\caption{\label{factor1a}Single particle survival probability and
bounds with $\rho_L = \rho_R = 0.5$.}
\end{figure}

First we consider the case of equal trap densities either side of the
origin and the case of $n=1$ and $2$ particles starting at the origin.
We generated the data for the case $n=1$ using the algorithm described
above, and densities $\rho_L=\rho_R=0.5$ until a time $t=30000$.
Bearing in mind the form of the bounds (\ref{genbounds}) it is
appropriate to plot the quantity $\lambda(t) = -\ln
Q(t)/\sqrt{\rho_-^2 Dt}$ against log time.  In all the simulations,
$D=D^\prime=\frac{1}{2}$ and in this case, $\rho_-=0.5$.  Hence the
upper and lower bounds in (\ref{genbounds}) converge to the constant
$\lambda(\infty) = 4/\sqrt{\pi}$.  Fig.~\ref{factor1a} shows that,
after an initial transient, $\lambda(t)$ does fall within the bounds.
However, even at the late times probed in the simulation, $\lambda(t)$
still seems to be far away from its asymptote.  This highlights the
fact that the predicted asymptotic form for the particle's survival
probability (\ref{BLresults}) has not yet been observed in simulation,
even with sophisticated methods at our disposal.

\begin{figure}
\includegraphics[width=\linewidth]{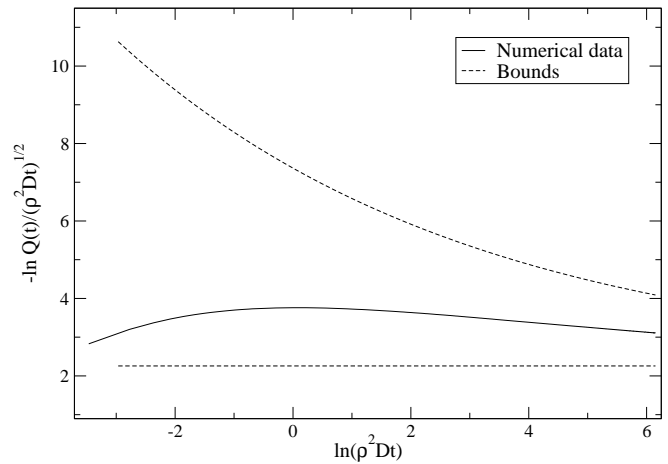}
\caption{\label{factor1b}Two particle survival probability taken from
\cite{MG02} and bounds with $\rho_L = \rho_R = 0.25$.}
\end{figure}

The data for the case $n=2$ have been taken from \cite{MG02} and are
plotted with our bounds in Fig.~\ref{factor1b}.  As with the case
$n=1$ we have from (\ref{genbounds}) that $\lambda(\infty) =
4/\sqrt{\pi}$ and again the convergence to asymptopia is very slow.

\begin{figure}
\includegraphics[width=\linewidth]{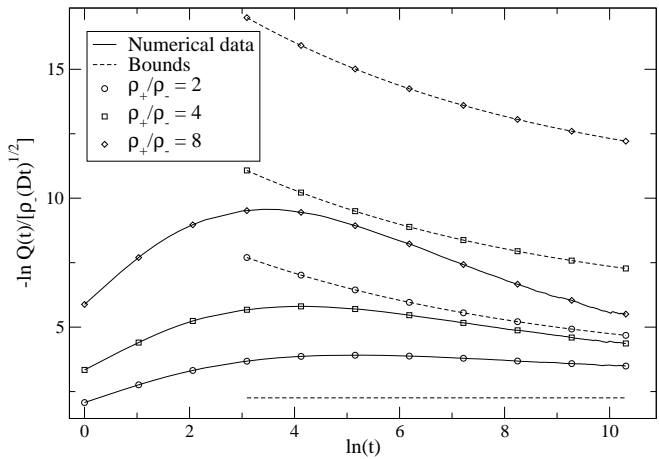}
\caption{\label{factor248}Single particle survival probability and
bounds with $\rho_{+}/\rho_{-} = 2,4$ and $8$.  The symbols on the
solid lines (representing the numerical data) are included purely for
the purpose of identifying each curve with the corresponding density
ratio.}
\end{figure}

In Fig.~\ref{factor248} we plot the single particle survival
probability for the case where the densities of traps either side of
the origin are unequal.  Specifically we have the cases $\rho_+/\rho_-
= 2,4,8$ with $\rho_R=0.5$ in each case.  Note that the density used
to scale the plots is always the smaller of the two, $\rho_-$.  Again
we see that the numerical data lie within the bounds predicted by Eq.\
(\ref{genbounds}). In these cases, however, the bounds we have
presented do not converge so we have no predictions for the limiting
value of $\lambda(t)$.

\section{Analytical results in greater than one dimension}
\label{generaldanalysis}

The upper and lower bounds on $Q(t)$ derived in $d=1$ will be now be
generalized to all $d$ in the range $1<d<2$ and to $d=2$, the latter
case requiring a slightly different treatment.  The case $d>2$ will
also be discussed.

\subsection{Upper Bound}
Let the particle, with diffusion constant $D'$, start at the origin, 
and the traps, with diffusion constant $D$, be randomly distributed 
in space with density $\rho$. As before, we assert, on intuitive 
grounds, that the ``best strategy'' for the particle is to stay 
at rest at the origin. With this assumption (which was verified 
numerically for $d=2$, for all times up to $t=12$, in the preceding 
section) the survival probability for $D'=0$ provides an upper bound 
on the survival probability for any $D'>0$. Let $Q_1(t|r)$ be the 
probability that a given trap, starting a distance $r$ from the origin, 
has not yet visited the origin at time $t$. It obeys the backward 
Fokker-Planck equation 
\begin{eqnarray}
\frac{\partial Q_1}{\partial t}  & = & D\nabla^2 Q_1 \nonumber \\
& = & D\left(\frac{\partial^2 Q_1}{\partial r^2} + \frac{d-1}{r}\,
\frac{\partial Q_1}{\partial r}\right)\ ,
\label{BFPE}
\end{eqnarray}
where we have exploited the spherical symmetry of the problem. 
The boundary conditions are $Q_1(t|0) = 0$ for all $t$ and 
$Q_1(t|\infty) = 1$ for all $t$, while the initial condition is 
$Q_1(0|r) = 1$ for all $r>0$. Since there is no length scale in the 
problem, $Q_1(t|r)$ must have the scaling form 
\begin{equation}
Q_1(t|r) = f(r/\sqrt{Dt})\ . 
\label{scaling}
\end{equation}
Substituting this form into Eq.\ (\ref{BFPE}) gives an ordinary 
differential equation for $f(x)$: 
\begin{equation}
\frac{\D^2 f}{\D x^2} + \frac{d-1}{x}\,\frac{\D f}{\D x} + \frac{x}{2}
\frac{\D f}{\D x} = 0\ ,
\end{equation}
with boundary conditions $f(0)=0$, $f(\infty) = 1$. The solution is 
\begin{equation}
f(x) = \left[\Gamma\left(\frac{2-d}{2}\right)\right]^{-1}\,
\int_0^{x^2/4} \Di{s} s^{-d/2}\,e^{-s}\ .
\label{f}
\end{equation} 
For $d=1$ our previous result, $f(x) = \erf(x/2)$, is recovered. 
Note that Eq.\ (\ref{f}) is only valid for $d<2$, since the integral 
diverges for $d \ge 2$. This regime will therefore require a different 
treatment. 

Eq.\ (\ref{f}) gives the survival probability of a stationary particle in 
the presence of a single diffusing trap. Consider $N$ traps in a large 
sphere of volume $V$ centered on the origin. Each trap starts anywhere 
in the volume with equal probability. The average, over the initial 
positions of the traps, of the probability that none of the traps has 
yet reached the origin at time $t$ is 
\begin{eqnarray}
Q(t) & = & \left[\frac{1}{V}\int_V \D^dr f\left(\frac{r}{\sqrt{Dt}}\right)
       \right]^N \nonumber \\
& = & \left[1 -\frac{1}{V}\int_V \D^dr 
\left\{1 - f\left(\frac{r}{\sqrt{Dt}}\right)\right\}\right]^N . 
\end{eqnarray}
Taking the limit $N \to \infty$, $V \to \infty$, with $\rho=N/V$ 
held fixed, gives
\begin{equation}
Q(t) = \exp\left[ -\rho\int \D^dr 
\left\{1 - f\left(\frac{r}{\sqrt{Dt}}\right)\right\}\right]\ ,
\label{Qinfinity}
\end{equation} 
where the integral is now over all space. Inserting the function 
$f(x)$ from Eq.\ (\ref{f}) and evaluating the integral gives the 
final result, which serves as an upper bound, $Q_U(t)$, for the 
problem with general $D'>0$: 
\begin{equation}
Q_U(t) = \exp[-a_d \rho (Dt)^{d/2}]\ ,
\label{upper}
\end{equation}
where
\begin{equation}
a_d = \frac{2}{\pi d}\,(4\pi)^{d/2}\,\sin\left(\frac{\pi d}{2}\right)\ .
\label{a_d}
\end{equation}

\subsection{Lower Bound}
Our strategy for constructing a rigorous lower bound follows that
employed in one dimension. We construct an imaginary ($d$-dimensional)
sphere of radius $l$ centered on the origin, and calculate the
probability that (i) there are no traps inside the sphere at $t=0$
(ii) the particle stays inside the sphere up to time $t$, and (iii) no
traps enter the sphere up to time $t$. As before, the set of
trajectories (of particle and traps) selected by these constraints are
a subset of all trajectories in which no traps meet the particle, so
the probability weight of this subset provides a lower bound on
$Q(t)$. We compute these probabilities in turn.

(i) The probability that the sphere initially contains no traps is 
simply $\exp(-\rho V_d l^d)$, where $V_d = 2\pi^{d/2}/d\Gamma(d/2)$ 
is the volume of a $d$-dimensional unit sphere. 

(ii) The probability, $Q_P(t|r,l)$ that the particle stays inside the 
sphere up to time $t$ is obtained by solving the backward Fokker-Planck 
Eq.\ (\ref{BFPE}), with $D$ replaced by $D'$, subject to the boundary 
conditions $Q_P(t|l,l)=0$ and $Q_P(t|r,l)$ is analytic at $r=0$, and 
the initial condition $Q_P(0|r,l)=1$ for $r<l$. The solution has the form 
\begin{equation}
Q_P(t|r,l) = r^\nu \sum_{n=1}^\infty c_n \exp(-D'k_n^2t)J_{-\nu}(k_nr)\ ,
\end{equation}
where $J_{-\nu}(z)$ is a Bessel function of the first kind, 
\begin{equation}
\nu = (2-d)/2\ ,
\end{equation}
and $k_nl = z_n$ is the $n$th zero of $J_{-\nu}(z)$. The coefficients 
$c_n$ are obtained from the initial condition, but their precise values 
are of no interest here. Since the particle starts at $r=0$, we need 
$Q_P(t|0,l)$. Its asymptotic form is 
\begin{equation}
Q_P(t|0,l) \sim \exp(-z_1^2D't/l^2)\ .
\end{equation}  

(iii) To compute the probability, $Q_T(t)$, that no trap enters the sphere 
up to time $t$ (the target annihilation problem) we begin by calculating 
this probability, $Q_1(t|r,l)$,  for a single trap. Then the probability 
that none of the traps enter the sphere is given by a natural generalization 
of Eq.\ (\ref{Qinfinity}),
\begin{equation}
Q_T(t) = \exp\left[-\rho\int_{r>l} \D^dr 
\left\{1 - Q_1(t|r,l)\right\}\right]\ .
\label{Qinfinity1}
\end{equation}  
In contrast to the case $l=0$ used for the upper bound, there is no 
simple scaling form analogous to (\ref{scaling}) for $Q_1(t|r,l)$ 
because $l$ provides an additional length scale. 

The function $Q_1(t|r,l)$ obeys the backward Fokker-Plank equation
(\ref{BFPE}), with boundary conditions $Q_1(t|l,l) = 0$ for all $t$, 
$Q_1(t|\infty,l) = 1$ for all $t$, and initial condition 
$Q_1(0|r,l)=1$ for $r > l$. The solution can be found by Laplace 
transform techniques. The result is \cite{BT93}
\begin{equation}
Q_1(t|r,l) = \frac{2}{\pi}\left(\frac{r}{l}\right)^\nu\int_0^\infty 
\frac{\D k}{k}\exp(-Dk^2t)G_\nu(kr,kl),
\label{Qinfinity2}
\end{equation}
where
\begin{equation}
G_\nu(x,y) =  \frac{Y_\nu(x)J_\nu(y)-J_\nu(x)Y_\nu(y)}{J^2_\nu(y) 
+ Y^2_\nu(y)}\ ,
\end{equation}
and $Y_\nu(z)$ is a Bessel functions of the second kind. 

Before continuing, we can first simplify Eq.\ (\ref{Qinfinity1}) as 
follows. First define 
\begin{equation} 
F(t) = \int_{r>l}\D^dr\,\left\{1 - Q_1(t|r,l)\right\}\ , 
\end{equation} 
where $F(0)=0$ follows from the initial condition $Q_1(0|r,l)=1$ for 
all $r>l$. Then we use the backward Fokker-Planck equation (\ref{BFPE}) 
to write
\begin{eqnarray}
\partial_t F & = & -D \int_{r>l} \D^dr\,\nabla^2 Q_1(t|r,l) \nonumber \\
 & = & -D \int_A {\bf dA} \cdot \nabla Q_1(t|l,l) \nonumber \\
\label{surfint}
 & = & D  S_d l^{d-1} \partial_r Q_1(t|r,l)|_{r=l},
\end{eqnarray}
where $A$ is the surface of the sphere, ${\bf dA}$ is a surface 
element directed along the inward normal to the sphere, and  
\begin{equation}
S_d = \frac{2 \pi^{d/2}}{\Gamma(d/2)}
\end{equation}
is the surface area of the unit sphere in $d$ dimensions. 
Integrating the result (\ref{surfint}) with respect to
time, with initial condition $F(0)=0$, Eq.\ (\ref{Qinfinity1}) 
takes the form \cite{BO87}
\begin{equation}
Q_T(t) = \exp\left[-\rho D S_d l^{d-1}\int_0^t \D t^\prime 
\partial_r Q_1(t^\prime|r,l)|_{r=l}\right]\ .
\label{Qinfinity3}
\end{equation}  
We are interested in the behavior of $Q_T(t)$ for large $t$. At this 
point it is convenient to discuss separately the cases $1<d<2$, $d=2$, 
and $d>2$.

\subsubsection{The case $1<d<2$}

For $1<d<2$, the function $Q_1(t|r,l)$, given by
Eq.\ (\ref{Qinfinity2}), has the large-$t$ expansion \cite{BT93}
\begin{eqnarray}
Q_1(t|r,l) & = & \left[\left(\frac{r}{l}\right)^{2\nu}-1\right]\,
\left\{\frac{\tau^{-\nu}}{\Gamma(1+\nu)}\right. \nonumber \\
&& + \left.\frac{\tau^{-2\nu}}{\Gamma^2(1+\nu)}
\frac{\Gamma^2(1-\nu)}{\Gamma(1-2\nu)} + \cdots\right\}, 
\end{eqnarray}
where $\tau = 4Dt/l^2$ and we recall that $\nu = (2-d)/2$. 
Taking the derivative with respect to $r$, setting $r=l$, 
inserting the result into Eq.\ (\ref{Qinfinity3}), and 
evaluating the integrals over $t'$, gives the probability, 
that the target has not been annihilated by a trap, 
\begin{equation}
Q_T(t) = \exp[-a_d\rho(Dt)^{d/2} - b_d\rho(Dt)^{d-1}l^{2-d} + \cdots],
\end{equation} 
where $a_d$ is given by Eq.\ (\ref{a_d}) and 
\begin{equation}
b_d = \frac{2^{2d-1}\pi^{d/2}\Gamma(d/2)}{(2-d)\Gamma^2(1-d/2)\Gamma(d)}\ .
\label{b_d}
\end{equation}
Note that, as with the case $d=1$, the leading term is independent of
$l$.  This phenomenon can be attributed to the recurrence of diffusion
in dimensions $d<2$.

Finally we assemble the contributions (i)--(iii) above to obtain a  
rigorous lower bound on the asymptotic behavior for $1<d<2$, 
\begin{eqnarray}
Q_L(t) & \sim & \exp[-a_d\rho(Dt)^{d/2}] \times\exp[-\rho V_dl^d  
\nonumber \\
&& - z_1^2D't/l^2 - b_d\rho(Dt)^{d-1}l^{2-d}]. 
\label{QL}
\end{eqnarray} 
As usual, for a given time $t^\ast$ we choose a sphere radius $l^\ast$
to optimize the lower bound.  The dominant $l$-dependent terms for $t
\to \infty$ are the final two terms in the second
exponential. Ignoring constants of order unity, we find that the value
of $l^\ast$ that gives the greatest lower bound is
\begin{equation}
l^\ast \sim \left(\frac{D'}{\rho
D^{d-1}}\right)^{1/(4-d)}(t^\ast)^{(2-d)/(4-d)} \;.
\end{equation}      
Inserting this into (\ref{QL}) the second exponential takes the form 
\begin{equation}
\exp[-{\rm const}(D')^{(2-d)/(4-d)}(\rho D^{d-1})^{2/(4-d)}t^{d/(4-d)}].
\end{equation} 
The neglected first term in the second exponential in (\ref{QL}) behaves 
as $l^d \sim t^{d(2-d)/(4-d)}$, which is indeed negligible compared 
to $t^{d/(4-d)}$ for large $t$ (recalling that $d>1$ here). 

In summary, the best lower bound behaves as
\begin{equation}
Q_L(t) \sim \exp[-a_d \rho (Dt)^{d/2} + \Order(t^{d/(4-d)})].
\label{QL2}
\end{equation}
Since $d/(4-d)<d/2$ for $d<2$, the two bounds pinch asymptotically, to 
give the exact result
\begin{equation}
\lim_{t \to \infty} -\frac{\ln Q(t)}{\rho(Dt)^{d/2}} = a_d, \ 1 \le d<2,  
\end{equation}
where we recall that $a_d$ is given be Eq.\ (\ref{a_d}). The constant 
$\lambda_d$ in Eq.\ (\ref{BLresults}) is therefore given by 
\begin{equation}
\lambda_d = \frac{2\rho}{\pi d}\sin\left(\frac{\pi d}{2}\right)\, 
(4\pi D)^{d/2}\ , \ \ \ \ 1 \le d < 2 \ .
\label{lambdad}
\end{equation}

Note that the subdominant term in (\ref{QL2}) decays more slowly 
relative to the leading term as $d \to 2$, signaling a change of 
behavior at $d=2$. Note also that the coefficient $a_d$ vanishes 
at $d=2$, suggesting a {\em slower} decay than a simple exponential 
in two dimensions. We now show that this expectation is correct, and 
determine the constant $\lambda_2$ in Eq.\ (\ref{BLresults}).

\subsubsection{The case $d=2$}

For $d=2$ the asymptotic form of $Q_1(t|r,l)$ is \cite{BT93}
\begin{equation}
Q_1(t|r,l) = 2\ln\left(\frac{r}{l}\right)\,\left[\frac{1}{\ln \tau} 
+ \Order\left(\frac{1}{\ln^2\tau}\right)\right]\ ,
\end{equation}
where $\tau = 4Dt/l^2$ as before. Inserting this into
Eq.\ (\ref{Qinfinity3}), with $d=2$, gives the probability that no trap
has entered the circle of radius $l$ up to time $t$:
\begin{equation}
Q_T(t) = \exp\left[-\frac{4\pi\rho D t}{\ln(4Dt/l^2)} 
+ \Order\left(\frac{t}{\ln^2 t}\right)\right]\ . 
\label{target2d}
\end{equation}
Following our previous procedure, the asymptotic lower bound is given by 
\begin{eqnarray}
Q_L(t) & \sim & \exp\left[- \frac{4\pi\rho D t}{\ln(4Dt/l^2)} 
+ \Order\left(\frac{t}{\ln^2 t}\right)\right. \nonumber \\ 
&& \left. -\rho \pi l^2 - z_1^2 \frac{D't}{l^2} \right], 
\end{eqnarray}
where $z_1$ is now the smallest zero of $J_0(z)$. The dominant terms 
in the exponential for large $t$ are the first and last terms. 
Extremizing this bound with respect to $l$ at some fixed $t^\ast$ gives 
\begin{equation}
l^\ast \sim z_1\left(\frac{D'}{4\pi\rho D}\right)^{1/2}
\ln\left(\frac{\rho D^2t^\ast}{D'}\right) 
\end{equation}
to leading order, and
\begin{equation} 
Q_L(t) \sim \exp\left[-\frac{4\pi\rho Dt}{\ln(\rho D^2t/D')} 
+ \Order\left(\frac{t\ln(\ln t)}{\ln^2 t}\right)\right]\ .
\label{2dlower}
\end{equation}

As far as the upper bound is concerned, Eq.\ (\ref{upper}) is   
not useful in $d=2$, since $a_2=0$. This tells us that the probability 
that a trap will reach a specified region of zero volume (i.e.\ a 
specified point) is zero in two dimensions. The $A$ particle has to be 
given a non-zero size (or the system put on a lattice) for a non-zero 
trapping probability. We therefore assign the particle a non-zero radius   
$a$, but still treat it as stationary for the upper bound. The traps will, 
for the moment, continue to be treated as point particles. With the 
definition that trapping occurs if a trap enters within the particle's 
radius (so that $a$ is an interaction range), our upper 
bound is just given by the probability $Q_T(t)$, Eq.\ (\ref{target2d}), 
but with $l$ replaced by $a$, to give
\begin{equation}
Q_U(t) \sim  \exp\left[-\frac{4\pi\rho D t}{\ln(4Dt/a^2)} 
+ \Order\left(\frac{t}{\ln^2 t}\right)\right]\ . 
\label{2dupper}
\end{equation}
In the limit $t\ \to \infty$, the bounds converge to give the 
asymptotic result $Q(t) \sim \exp(-4\pi\rho Dt/\ln t)$ or, equivalently 
\begin{equation}
\label{2dasymp}
\lim_{t \to \infty} - \frac{\ln t \ln Q(t)}{\rho D t} = 4\pi\ .
\end{equation}
This gives the constant $\lambda_2$ in Eq.\ (\ref{BLresults}) as 
\begin{equation}
\label{lambda2}
\lambda_2 = 4\pi\rho D\ .
\end{equation}

\begin{figure}
\includegraphics[width=\linewidth]{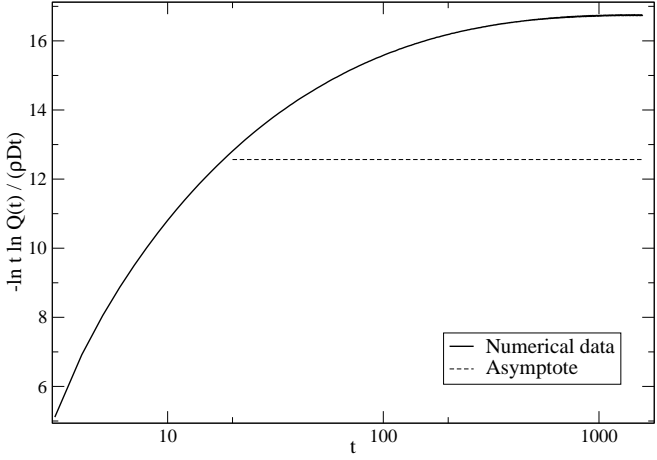}
\caption{\label{twod}Numerical data for the two-dimensional trapping
reaction taken from \cite{MG02}.  In the simulation, the trap density
$\rho = 1/4$ and the diffusion constants $D=D^\prime=1/4$. The
asymptote given by Eq.\ (\ref{2dasymp}) is plotted for comparison.}
\end{figure}

As noted previously, the algorithm described in Section~\ref{numerics}
can be used to simulate the trapping reaction in any integer
dimension.  Numerical results for the two-dimensional system were
presented in \cite{MG02} and we compare these data with the asymptotic
result (\ref{2dasymp}) in Fig.~\ref{twod}.  We find that the deviation
of the numerical results from the asymptote is even more marked in two
dimensions than in one (see Fig.~\ref{factor1a}). Part of the reason
for this is, presumably, that the increased number of sites in two
dimensions means that one cannot probe such late times. A second, and
perhaps more important, reason is the very large corrections to
scaling evident from our bounding arguments. The {\em relative} size
of the subleading term in Eq.\ (\ref{2dupper}) is $\Order(1/\ln t)$,
while the subleading term for the lower bound, Eq.\ (\ref{2dlower}),
is even larger, of relative size $\Order(\ln[\ln t]/\ln t)$. This
suggests that convergence to asymptopia will be \textit{extremely}
slow in two dimensions. Note also that the particle's survival
probability was found to decay to $\sim 10^{-99}$ after $t=1600$ time
steps.  This emphasizes the importance of determining the corrections
to asymptopia in order to determine the form of the survival
probability in numerically (and, indeed, experimentally) accessible
regimes.

\subsubsection{The case $d>2$}

The same bounding arguments can be applied equally well in $d>2$.  The
main difference from $d \le 2$ is that the bounds no longer converge,
so it is not possible to determine $\lambda_d$ exactly (except for $d$
very close to 2---see below). The basic idea is the same as for $d \le
2$, except that the particles must be given non-zero sizes (or,
equivalently, a non-zero range of interaction). We let the particle
have radius $a$, and the traps radius $b$.  A reaction is deemed to
have occurred if there is an overlap between the particle and any
trap, i.e.\ if the centers approach more closely than a distance
$R=a+b$, which is the range of interaction. (Note, however, that we
continue to assume that the traps do not interact with each other. In
particular, there is no excluded volume interaction between traps.)

The upper bound is obtained from the target annihilation problem 
with target radius $R$. For $d>2$, the single-trap  survival
probability, $Q_1(t|r,R)$, has a non-vanishing large-$t$ limit  
given by the well-known result
\begin{equation} 
Q_1(\infty|r,R) = 1 - (R/r)^{d-2}\ ,
\end{equation}
which is easily obtained from Eq.\ (\ref{BFPE}) on setting the left 
side to zero, and imposing the boundary conditions $Q_1(\infty|R,R)=0$, 
$Q_1(\infty|\infty,R)=1$ on the resulting ordinary differential 
equation. Inserting this form in Eq.\ (\ref{Qinfinity3}), with $l=R$, and 
evaluating the time integral, gives an upper bound with the leading large-$t$ 
behavior
\begin{equation}
Q_U(t) \sim \exp[-(d-2)S_d \rho R^{d-2} Dt]\ .
\label{d>2upper}
\end{equation}

The lower bound is obtained in a similar fashion, following the
pattern established for $d \le 2$. One constructs a notional sphere of
radius $l$, centered on the initial position of the particle. The
bound is given by the subset of trajectories in which (i) there are no
traps initially within the sphere, (ii) the center of the particle
remains within a sphere of radius $l-a$, so that the particle remains
entirely inside the sphere of radius $l$, and (iii) the center of
every trap remains outside a sphere of radius $l+b$, so that every
trap remains entirely outside the sphere of radius $l$.

The probability of (i) is $\exp[-\rho V_d (l+b)^d]$. The probability of (ii) 
has the asymptotic form $\exp[-z_1^2 D't/(l-a)^2]$, where $z_1$ is the first 
zero of $J_{-\nu}(z) = J_{(d-2)/2}(z)$. The probability of (iii) is given, 
for large $t$, by Eq.\ (\ref{d>2upper}) with $R$ replaced by $l+b$. 
Assembling these three contributions gives the asymptotic lower bound 
\begin{eqnarray}     
Q_L(t) & \sim & \exp[-(d-2)S_d \rho (l+b)^{d-2} Dt - z_1^2D't/(l-a)^2 
 \nonumber \\
&& \hspace*{4cm} - \rho V_d (l+b)^d]\ .
\end{eqnarray}
This has to be maximized with respect to $l$. For $t \to \infty$, the first 
two terms in the exponent dominate, and the final term is negligible.  
Setting $l=a+x$, and maximizing with respect to $x$, gives the equation 
\begin{equation}
(d-2)^2S_d\rho D(x+R)^{d-3} = 2z_1^2 D'/x^3\ ,
\label{x}
\end{equation}  
where $R=a+b$ as before. This equation cannot be solved analytically for 
general $d$, so we concentrate on two soluble cases---the physically 
interesting case $d=3$, and the limit $d \to 2+$. 

For $d=3$, we have $S_d = 4\pi$ and $z_1=\pi$, giving 
$x = (\pi D'/2\rho D)^{1/3}$ and 
\begin{equation}
Q_L(t) \sim \exp[-4\pi \rho DRt - 3(2\pi^2 \rho D\sqrt{D'})^{2/3}t]\ .
\label{3dlower}
\end{equation}
Combining the two bounds, we obtain the asymptotic form  $Q(t) \sim 
\exp(-\lambda_3 t)$, as in Eq.\ (\ref{BLresults}), with the bounds 
\begin{equation}
4\pi\rho DR \le \lambda_3 \le 4\pi \rho DR 
+ 3(2\pi^2 \rho D\sqrt{D'})^{2/3}\ ,\ \ d=3\ .
\label{lambda3}
\end{equation} 
It is worth noting that the second term on the right is negligible 
compared to the first if $D'/D \ll \rho R^3$, i.e.\ when $D'/D$ is 
small compared to the number of traps per interaction volume. 

For $d = 2+ \epsilon$, Eq.\ (\ref{x}) has the solution 
\begin{equation}
x = \frac{1}{\epsilon} \left(\frac{z_1^2 D'}{\pi\rho D}\right)^{1/2}\ ,
\label{xeps}
\end{equation}
to leading order for $\epsilon \to 0$, giving the lower bound 
$Q_L(t) \sim \exp[-2\pi\epsilon\rho Dt]$ to leading order in 
$\epsilon$. In the same limit, the upper bound (\ref{d>2upper}) has 
exactly the same form, giving the result $Q(t) \sim \exp(-\lambda_d t)$ 
with 
\begin{equation}
\lambda_d = 2\pi\rho D\, \epsilon + \cdots\ .
\label{eps}
\end{equation} 
Hence the bounds pinch to leading order in $\epsilon$, but not for 
general $d$.

To conclude this section, we consider again the case where $n$ particles 
start from the origin, and we want the probability that {\em all} survive 
until time $t$. As noted in the discussion of the one-dimensional case, 
$n$ only enters in the calculation of the lower bound, in the term 
giving the probability for the particle to stay inside the notional 
box ($d=1$), or sphere ($d>1$), of size $l$. This probability behaves as 
$\exp(-{\rm const} D't/l^2)$, so having $n$ particles simply requires 
raising this factor to the power $n$, which is equivalent to replacing 
$D'$ by $nD'$. Since the asymptotic forms we derive do not depend on $D'$ 
for $d \le 2$, it follows that our results are independent of $n$ in this 
regime. For $d>2$, however, our results do depend on $D'$ (see Eq.\ 
(\ref{lambda3})). In this regime, therefore, the generalization to 
arbitrary $n$ is achieved through the replacement $D' \to nD'$.  

\section{Discussion and Summary}
\label{summary}
In this paper we have derived a number of results for the asymptotic
survival probability of a particle diffusing among randomly
distributed diffusing traps with density $\rho$. We allow the particle
and traps to have different diffusion constants, $D'$ and $D$
respectively. Our results take the forms originally derived by Bramson
and Lebowitz \cite{BL88}, as expressed in Eq.\ (\ref{BLresults}).
With one assumption, supported by numerical evidence, we have obtained
exact results for the coefficients $\lambda_d$ in (\ref{BLresults})
for dimensions $d \le 2$, and an exact inequality for dimensions
$d>2$.  These results are given by Eqs.\ (\ref{lambda1}),
(\ref{lambdad}), and (\ref{lambda3}). For $d \le 2$ the results for
$\lambda_d$ are independent of the diffusion constant $D'$ of the
particle.

The results are obtained by deriving upper and lower bounds for
$\lambda_d$, and showing these coincide for $d \le 2$.  Whilst our
lower bound is rigorous, we had to assume that the particle's survival
probability for $D'=0$ provides an upper bound on its survival
probability when $D'>0$ when the trap distribution is symmetric.
Indeed, for the $d=1$ system with different densities of traps to the
left and right of the particle, it was found that staying still is
{\em not} the particle's best strategy. Instead, trajectories that
survive for long times tend to be those in which the particle drifts
to the side with the lower trap density. This emphasizes the crucial
role of the symmetry of the trap distribution, an observation
supported by perturbative studies for a system with a finite number of
traps \cite{BB02a,BB02b}.

In all cases the particle and traps are assumed to move in a 
continuous space, and to have zero size for $d<2$. For $d \ge 2$ 
is is necessary for the particle and/or the traps to have non-zero 
size (otherwise the survival probability, for motion on a 
continuous space, is one for all time). We also take the traps 
to be randomly distributed in space at time $t=0$, with uniform 
density $\rho$. This raises the question of the extent to which 
the results are ``universal'', i.e.\ independent of the microscopic 
details of the model, a question which we now address. 

We argue that, for $d \le 2$, the results do indeed have a 
degree of universality. In $d=1$, the optimal box size used to obtain
the lower bound on $Q(t)$ is {\em large}, $l \sim t^{1/3}$, as $t \to
\infty$, so the effect of the particle having a finite size when
confined to this box is negligible.  With a little thought one sees 
that the same is true for all $d \le 2$. The optimal length scale 
for the lower bound {\em grows with time} as $l \sim t^{(2-d)/(4-d)}$ 
($d < 2$) or $l \sim \ln t$ (d=2), and the results are independent 
of the particle and trap sizes, as far as the leading-order results 
are concerned. The same is true of the upper bound---the finite-size 
corrections come in at subleading order. 

The dominance of large length scales at late times, for $d \le 2$, 
also suggests that the asymptotic results are independent of whether 
the model is defined on the continuum (as here) or on a lattice, an 
assumption implicitly made earlier when we compared our theoretical 
predictions to numerical results obtained from lattice simulations. 
For $d>2$, however the dominant value of $l$ that determines the lower 
bound is time-independent. Therefore we expect a lack of universality 
in this case. The explicit dependence on the interaction range $R$ in 
Eq.\ (\ref{lambda3}) is a signature of this effect. Note, however, that 
to leading order in $\epsilon=d-2$, $\lambda_d$ is independent of $R$ 
(see Eq.\ (\ref{eps})) and we expect the result to be universal to this 
order. Physically, this is because the length scale $l = a+x$ diverges 
as $\epsilon\to 0$ (see Eq.\ (\ref{xeps})). 

A further universality question concerns universality with respect to  
the initial conditions. We have taken Poissonian initial conditions, 
where the probability $P_N(V)$ of having $N$ traps in a volume $V$
is given by $P_N(V)= [(\rho V)^N/N!]\exp(-\rho V)$ for any $V$.
The lattice simulations, where the number of traps on each site has 
a Poisson distribution (with mean $\rho$, say) has this property, 
namely the number of traps on $m$ sites has a Poisson distribution 
with mean $m\rho$. Whether there is a larger class of initial conditions 
sharing the same asymptotic behavior is a question deserving further 
study. 

We conclude by discussing some recent papers related to the present
work, and directions for future work. The coefficient $\lambda_d$ in
Eq.\ (\ref{BLresults}) has recently been calculated using a
diagrammatic method \cite{vW02} to first order in $(2-d)$. The quoted
result, however, exceeds our rigorous upper bound for $\lambda_d$
(corresponding to the lower bound for $Q(t)$) by a factor of two.
This is because in \cite{vW02} $\lambda_d$ depends on $D$ and
$D^\prime$ only through their sum $D+D^\prime$ \cite{vWpc} whereas our
rigorous upper bound on $\lambda_d$ depends solely on $D$.  It is
interesting to note that in the related process $A+B\to A$ where the
single $A$ particle acts as a trap for the $B$ particles, certain
properties of the $B$-particle distribution \textit{can} be expressed
as functions of $D+D^\prime$ \cite{SRW98}.  However, we stress that
for the $A+B\to B$ reaction studied here, the asymptotics are entirely
governed by the $B$-particle diffusion constant for $d\le 2$.

In a very recent work \cite{Oshanin02} our approach, as outlined in
\cite{BB02a}, has been generalized to diffusion on fractals for the
case where the fractal dimension of the traps' trajectories is greater
that the physical dimension (this condition is the analog of the
condition $d<2$ in the present work). It should be noted, however,
that in Ref.\ \cite{Oshanin02} the optimal lower bound on $Q(t)$ is
not obtained.  For $d<2$, only the subdominant corrections to the
leading terms are affected, and the upper and lower bounds still pinch
asymptotically. For $d=2$, however, the approach used in
\cite{Oshanin02} yields bounds that no longer converge at large $t$,
so the exact result (\ref{lambda2}) for $\lambda_2$ is missed.

In this paper and our earlier work \cite{BB02a} we noted that the extant 
simulation data \cite{MG02} fail to reach the asymptotic regime even 
though survival probabilities are so small that they can only be measured 
using sophisticated methods.  Perhaps the most important challenge, 
therefore, is to obtain a better understanding of the corrections 
to asymptopia in order to make testable, quantitative predictions. 
Other directions for future work include exploring further the extent to 
which our results are universal, and establishing rigorously the 
validity of our upper bound. 

\begin{acknowledgments}
We thank F. van Wijland and G. Oshanin for discussions, and N. Agmon for 
drawing our attention to reference \cite{BT93}. 
R.A.B.\ thanks the EPSRC for financial support under grant GR/R53197.
\end{acknowledgments}

%\bibliographystyle{apsrev}
%\bibliography{trefs}
%
% bibTeX output, slightly hand-modified, follows.

\end{document}